\documentclass[usenatbib]{mn2e}
\pdfoutput=1
\usepackage{graphicx}
\usepackage{latexsym}
\usepackage{aas_macros}
\usepackage{amsfonts,amsmath,amssymb}
\usepackage{url}
\usepackage[utf8]{inputenc}
\usepackage{fancyref}

\title[The substellar companion in SDSS J1411+2009]{The substellar companion in the eclipsing white dwarf binary SDSS J141126.20+200911.1}

\author[S.\,P.\ Littlefair et al.]{S.\,P.\ Littlefair$^{1}$, S.\,L.\ Casewell$^{2}$, S.\,G.\ Parsons$^{3}$,  V.\,S.\ Dhillon$^{1}$, T.\,R.\ Marsh$^{4}$,  
\newauthor B.\,T.\ G\"{a}nsicke$^{4}$, S.\ Bloemen$^{5}$,  S.\ Catalan$^{4}$, P.\ Irawati$^{6}$, L.\,K.\ Hardy$^{1}$,  M.\ Mcallister$^{1}$,
\newauthor M.\,C.\,P.\ Bours$^{4}$, Andrea\ Richichi$^{6}$, M.\,R.\ Burleigh$^{2}$, B. Burningham$^{7}$,
\newauthor E. Breedt$^{4}$, P. Kerry$^{1}$\\
$^{1}$Dept of Physics and Astronomy, University of Sheffield, Sheffield, S3 7RH, UK \\
$^{2}$Department of Physics and Astronomy, University of Leicester, University Road, Leicester LE1 7RH, UK\\
$^{3}$Departmento de F\'{i}sica y Astronom\'{i}a, Universidad de Valpara\'{i}so, Avenida Gran Bretana 1111, Valpara\'{i}so 2360102, Chile\\
$^{4}$Dept of Physics, University of Warwick, Coventry, CV4 7AL, UK\\
$^{5}$Department of Astrophysics, IMAPP, Radboud University
Nijmegen, PO Box 9010, NL-6500 GL Nijmegen, Netherlands\\
$^{6}$National Astronomical Research Institute of Thailand, Chiang Mai, Thailand, 50200\\
$^{7}$Centre for Astrophysics Research, Science and Technology Research Institute, University of Hertfordshire, Hatfield AL10 9AB, UK}

\begin{document}



\maketitle

\label{firstpage}

\begin{abstract}
We present high time resolution SDSS-$g'$ and SDSS-$z'$ light curves of the primary eclipse in SDSS J141126.20+200911.1, together with time-resolved X-Shooter spectroscopy and near-infrared $JHK_{s}$ photometry. Our observations confirm the substellar nature of the companion, making SDSS J141126.20+200911.1 the first eclipsing white dwarf/brown dwarf binary known.  We measure a (white dwarf model dependent) mass and radius for the brown dwarf companion of $M_{2} = 0.050 \pm 0.002$\,$M_{\odot}$ and $R_{2} = 0.072 \pm 0.004$\,$M_{\odot}$, respectively. The lack of a robust detection of the companion light in the $z'$-band eclipse constrains the spectral type of the companion  to be later than L5. Comparing the NIR photometry to the expected white dwarf flux reveals a clear $K_s$-band excess, suggesting a spectral type in the range L7--T1. 

The radius measurement is consistent with the predictions of evolutionary models, and suggests a system age in excess of three Gyr. The low companion mass is inconsistent with the inferred spectral type of L7--T1, instead predicting a spectral type nearer T5. This indicates that irradiation of the companion in SDSS J1411 could be causing a significant temperature increase, at least on one hemisphere. 
\end{abstract}

\begin{keywords}
binaries: close - binaries: eclipsing - stars
\end{keywords}

\bibliographystyle{mn2e_fixed2}

\section{Introduction} 
The eclipsing white dwarf binary SDSS J141126.20+200911.1 (also known as CSS21055, hereafter called SDSS J1411) was first noted in the catalogue of \cite{drake10}, but independently discovered by \citet[][hereafter B13]{beuermann13}. The SDSS photometry and spectroscopy are dominated by the white dwarf component, and show no hint of a companion. From the SDSS data, \cite{kleinman13} derived an effective temperature of $T_{\rm eff} = 13074 \pm 1306$\,K, ${\rm log}\ g = 7.89 \pm 0.12$ and a mass of $M_{1} = 0.551$\,M$_{\odot}$, suggesting a carbon-oxygen core white dwarf. From white-light observations, B13 noted the narrow eclipse was indicative of a small companion. Bessel $I$-band observations with lower time resolution allowed B13 to create an in-eclipse $I$-band image, in which all the light comes from the companion. SDSS J1411 was not detected in this image, setting a  model-dependent 1$\sigma$ upper limit on the companion mass of 0.075$M_{\odot}$. The faintness of the companion in the optical thus suggests a substellar mass for the companion in SDSS J1411, which would make it the first eclipsing white dwarf/brown dwarf binary identified. 

Here we present photometric and spectroscopic observations of SDSS J1411.  We use these data to constrain the mass, radius and spectral type of the  companion  in SDSS J1411.  We confirm the substellar nature of the companion, firmly establishing SDSS J1411 as the first detached, eclipsing white dwarf/substellar binary identified. The observations are described in section~\ref{sec:observations}, the results are presented in section~\ref{sec:results}, and discussed in section~\ref{sec:discussion}.

\section{Observations}
\label{sec:observations}
\begin{table*}
\begin{center}
\begin{tabular}{lccccccccc}
\hline
Date & Start Phase & End Phase  & Instrument & $T_{mid}$  & Exp. time  &
Data points  & Seeing  & Airmass & Phot?\\ 
&&&& BMJD(TDB) & (seconds) &  &  (arcseconds)  & &  \\ \hline
2014 Jan 25 & 8179.941 & 8180.106  & USPEC & 56682.86665(2) & 6.53 & 124 & 1.6--3.1 & 1.20--1.22 & Yes \\
2014 Jan 25 & 8180.850 & 8181.131  & USPEC & 56882.95116(1) & 6.53 & 209 & 1.5--2.4 & 1.00--1.01 & Yes\\
2014 Jan 26 & 8191.931 & 8192.053  & USPEC & 56683.88102(1) & 6.53 & 92 & 1.4--1.6 & 1.14--1.18  & Yes \\
2014 Jan 27 & 8202.935 & 8203.046  & USPEC & 56684.81085(2) & 6.53 & 84 &  1.7--2.5 & 1.61--1.73 & Yes \\
2014 Jan 28 & 8214.904 & 8215.041  & USPEC & 56685.82528(2) & 6.53 & 103 & 1.7--3.4 & 1.44--1.55 & Yes\\
2014 Jan 29 & 8227.889 & 8228.045  & USPEC & 56686.92421(2) & 6.53 & 117 & 1.5--2.6 & 1.02--1.04 & Yes \\
2014 Jan 31 & 8251.945 & 8252.032  & USPEC & 56688.95300(2) & 8.39 & 66 & 2.0--3.1 & 1.00--1.01 & Yes \\
2014 Feb 01 & 8261.959 & 8262.067 & USPEC & 56689.79835(2) & 6.53 & 82 & 1.7--2.6 & 1.58--1.70 & Yes \\
2014 Mar 02 & 8607.828 & 8608.037 & UCAM & 56719.046632(6) & 2.04 & 741 & 1.4--3.4 & 1.09--1.14 & Yes \\
2014 Mar 16 & 8775.867 & 8776.125 & UCAM & 56733.248125(7) & 4.54 & 412 & 1.2--1.4 & 1.11--1.19 & No \\
2014 Mar 26 & 8887.909 & 8888.033 & USPEC & 56742.71576(2) & 9.89 & 93 & 1.4--1.6 & 1.18--1.23 & Yes\\
2014 Mar 26 & 8889.934 & 8890.048 & USPEC & 56742.88489(2) & 9.89 & 85 & 1.2--1.5 & 1.09--1.11 & Yes \\
2014 Mar 27 & 8899.939 & 8900.063 & USPEC & 56743.73021(2) & 9.89  & 93 & 1.3--1.5 & 1.10--1.14 & Yes\\
2014 Mar 27 & 8900.970 & 8901.075 & USPEC & 56743.81472(2) & 9.89 & 69 & 1.1--2.7 & 1.00--1.01 & Yes \\
2014 Mar 27 & 8901.958 & 8902.077 & USPEC & 56743.89923(2) & 9.89 & 89 & 1.2--1.4 & 1.25--1.05 & Yes\\
2014 Mar 28 & 8912.904 & 8913.080 & USPEC & 56744.82912(1) & 9.89 & 131 & 1.2--1.6 & 1.00--1.02 & Yes \\
2014 Mar 28 & 8913.966 & 8914.064 & USPEC & 56744.91365(2) & 9.89 & 73 & 1.2--1.7 & 1.23--1.27 & Yes\\
2014 Mar 30 & 8936.961 & 8937.090 & USPEC & 56746.85791(1) & 9.89 & 96 & 1.2--1.4 & 1.05--1.08  & Yes\\
2014 Mar 30 & 8937.847 & 8938.142 & USPEC & 56746.94244(2) & 9.989 & 216 & 1.3--2.4 & 1.38--1.62 & Yes\\
\hline
2014 Feb 20 & 8491.952 & 8492.600 & LIRIS/K & n/a   & 10.00 & 136 & Variable & 1.00-1.10 & Yes \\ 
2014 Mar 17 & 8787.693 & 8788.223 & LIRIS/H & n/a   & 10.00 & 125 & Variable & 1.10--1.30 & Yes \\ 
2014 Apr 08 & 9045.490 & 9045.834 & LIRIS/J & n/a    & 20.00 & 60   & Poor & 1.03--1.10 & No \\ 
\hline
2014 Apr 19 & 9165.047 & 9166.955 & X-Shooter & n/a & 450/600/300 & 28/21/44 & 0.7--1.0 & 1.65--1.66 & n/a \\ 
2014 Apr 20 & 9178.202 & 9178.753 & X-Shooter & n/a & 450/600/300 & 9/7/13 & 0.7--0.9 & 1.44--1.65 & n/a \\ 
\hline
\end{tabular}
\caption{\label{table:obs}Journal of observations. For ULTRACAM (UCAM) observations, the dead-time between exposures was 0.025~s, for ULTRASPEC (USPEC) the dead-time was 0.015~s. The relative GPS timestamping on each UCAM/USPEC data point is accurate to 50 $\mu$s, with an absolute accuracy better than 1 ms.  $T_{mid}$ gives the time of mid eclipse (see section~\ref{subsec:times}). For X-Shooter, we provide exposure times in the UVB/VIS and NIR arms respectively. The column Phot? indicates if the night was photometric.}
\end{center}
\end{table*}

\subsection{Optical Photometry}
We observed eclipses of SDSS J1411 on a number of nights between January and March 2014 with the fast photometer ULTRASPEC (Dhillon et al, 2014, MNRAS, submitted) on the 2.4-m Thai National Telescope (TNT) on Doi Inthanon, Thailand. Observations were carried out in the SDSS $z'$-band filter. In addition, we observed two eclipses of SDSS J1411 on the nights of  2014~Mar~$2^{nd}$ and 2014~Mar~$16^{th}$ with the triple-beam high-speed photometer ULTRACAM \citep{dhillon07} on the 4.2-m William Herschel Telescope (WHT) on La Palma. On these occasions, we observed simultaneously in the SDSS-$u'g'z'$ filters. A complete journal of observations is presented in Table~\ref{table:obs}. In addition to observations of the eclipses, we observed the field of SDSS J1411 once with each instrument, applying random offsets to the pointing, which allowed us to median-stack these images to create an image for removal of fringing from the $z'$-band images. 

Data reduction was carried out in a standard manner using the {\sc ultracam} pipeline reduction software, as described in \cite{feline04}. In addition to the usual application of bias and flat frames, the fringe frame was subtracted from each data frame after being scaled by a constant which was chosen to minimise the variance in the background of the fringe-subtracted data frame. ULTRASPEC on the TNT suffers from a significant ($\approx10$ percent) contribution from scattered light, thought to originate from inadequate baffling of the tertiary mirror. This scattered light also affects the structure in the fringe frame, and changes in the baffling following the observation of the fringe frame in January meant that fringe removal did not work for the data taken in March. Therefore, we created a fringe frame for the March data using the following recipe. $z'$-band images from a single night  (2014~Mar~26$^{th}$) were combined to make a deep image in which stars were detected using {\sc sextractor}. Each star in this catalogue was removed from each individual $z'$-band image by fitting and subtracting a Moffat-profile \citep{trujillo2001} to the star. The star-subtracted images were then median stacked to produce a fringe frame.  The lightcurves of three nearby stars were averaged to produce a comparison star lightcurve in order to correct the data for transparency variations. The effective SDSS $z'$-band magnitude of this comparison light curve was $z' = 14.859$.

\subsection{Infrared Photometry}

The near-IR data in the $J$, $H$, and $K_s$ bands were obtained using LIRIS on the WHT as part of service programme SW2013b38. The data were reduced using \textsc{iraf}. The images were flat-fielded and bad pixel corrected before the \textsc{mosaic} task (part of package \textsc{xdimsum}) was used to create the sky background image, which was subtracted from each science frame, before averaging all images to create a final, stacked image. We then performed aperture photometry on the stacked images in each band using \textsc{sextractor} in the \textsc{starlink gaia}  package and an aperture equal to 1.5 times the image full-width half-maximum. For each  image, five reference stars with magnitudes listed in the 2MASS catalogue were used to derive a zeropoint, which was then used to calibrate the photometry. As both LIRIS and 2MASS use the same filter set, no additional colour corrections were applied.

\subsection{X-shooter Spectroscopy}
We observed SDSS J1411 with the medium resolution spectrograph X-shooter \citep{dodorico06} mounted at the Cassegrain focus of VLT-UT2 at Paranal. X-shooter is comprised of three detectors: the UVB arm, which gives spectra from 0.3-0.56 microns, the VIS arm which covers 0.56-1 microns and the NIR arm which covers 1-2.4 microns. We used slit widths of 1.0, 0.9 and 0.9 arcseconds in X-shooter's three arms and binned by a factor of two in the dispersion direction in the UVB and VIS arms, and a factor of two in the spatial direction in the UVB arm, resulting in a spectral resolution of 2500--3500 across the entire spectral range.

The reduction of the raw frames was conducted using the standard pipeline release of the X-shooter Common Pipeline Library (CPL) recipes (version 6.4.1) within ESORex, the ESO Recipe Execution Tool, version 3.10.2. The standard recipes were used to optimally extract and wavelength calibrate each spectrum. The instrumental response was removed by observing the spectrophotometric standard star LTT\,3218 and dividing it by a flux table of the same star to produce the response function. The wavelength scale was also heliocentrically corrected. A telluric correction was applied using observations of the DQ white dwarf GJ\,440 obtained just before the start of our observations and implemented using the \textsc{Molecfit} package \citep{kausch14}.

X-shooter is ideally suited for studying the composite spectra of post common envelope binaries since its long wavelength range potentially provides features from both stars simultaneously, allowing precision measurements to be made (see for example \citealt{parsons12a,parsons12b,parsons13}).  We searched the X-shooter spectrum for features from the substellar component but no emission line or absorption line features attributable to the companion were found. However, the NIR spectra have very low signal-to-noise hence any features at these wavelengths may not have been picked up. In our subsequent analysis we ignore the NIR spectra.

\section{Results}
\label{sec:results}

\subsection{The White Dwarf in SDSS J1411}
\label{subsec:wd}
\begin{figure*}
\begin{center}
\includegraphics[width=2.0\columnwidth,trim=90 10 90 10]{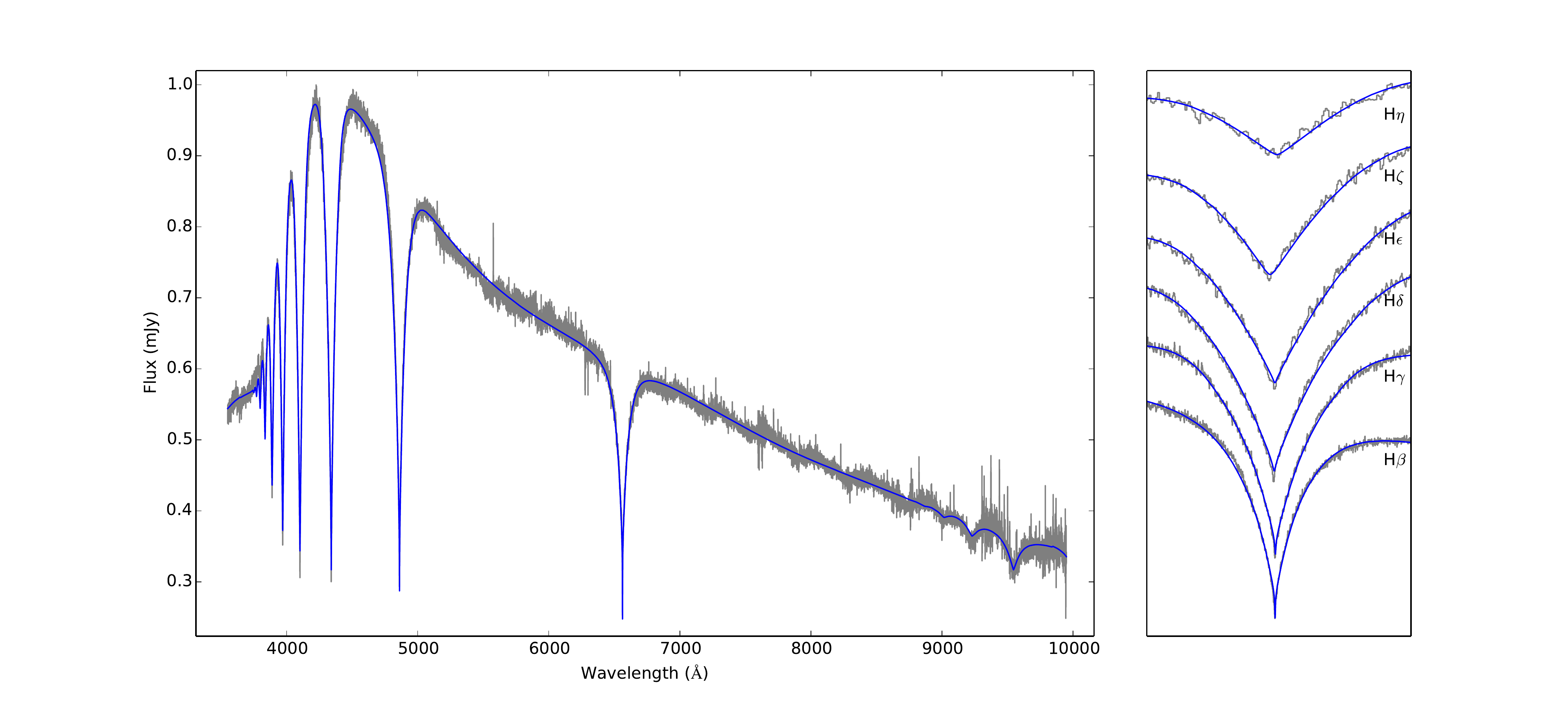}
\caption{\label{fig:av_wd_spec} {\bf Left: }The average X-Shooter spectrum, and best fit model. See section~\protect\ref{subsec:wd} for details of model. The model has been multiplied by a 5$^{th}$-order polynomial to account for overall normalisation, unknown reddening and uncertainties in flux calibration of the X-Shooter spectrum. The best fit model has $\chi^{2} = 65436$ with 25996 degrees of freedom. {\bf Right:} a zoom-in on the higher order Balmer lines.}
\end{center}
\end{figure*}
In order to find the effective temperature and gravity of the white dwarf, we fit the average X-Shooter spectrum, shifted into the reference frame of the white dwarf (see section~\ref{subsec:rv}). Our model consists of a set of DA white dwarf model spectra with mixing length ${\rm ML2}/\alpha = 0.8$ \citep{koester10} computed on a grid of $7.25 \le \log g \le 8.5$ in steps of 0.25 dex and $12000 \le T_{\rm eff} \le13750$\,K in steps of 250\,K. Linear bivariate interpolation was used to calculate model spectra at a given value of $\log g$ and $T_{\rm eff}$, and synthetic spectra produced by computing the average of three points within each spectral bin of the X-Shooter spectrum. To account for the effects of extinction and uncertainties in flux calibration of the X-Shooter spectrum, we follow \cite{eisenstein06} and marginalise over a polynomial of order five that multiplies the model spectrum. This ensures our best fit values are controlled by the line widths and strengths, rather than the continuum shape. 

We estimated the posterior probability distributions for $\log g$ and $T_{\rm eff}$ using an implementation of the Affine-Invariant Markov-Chain Monte-Carlo (MCMC) sampler \citep{foreman-mackey13}.  A systematic error on the flux values was included as a 'nuisance parameter' in the model.  We estimate the most likely values for each parameter by the median of all values in the MCMC chains. This suggests a white dwarf temperature of $T_{\rm eff} = 13000 \pm 15$\,K, and $\log g = 7.86 \pm 0.01$. The X-Shooter spectrum, and most likely model, are shown in Figure~\ref{fig:av_wd_spec}. The uncertainties quoted above are purely statistical and do not account for systematic uncertainties either in the spectroscopic data, models or fitting procedure. We follow \cite{napiwotzki99} and allow systematic errors of 2.3 percent in $T_{\rm eff}$ and 0.07 dex in $\log g$, to obtain final estimates of $T_{\rm eff} = 13000 \pm 300$\,K, and $\log g = 7.86 \pm 0.07$. Our results are compatible with, but more precise than, the results of \cite{kleinman13}.

Additional constraints on the white dwarf can be found using the SDSS photometry and the cooling models of \cite{holberg2006}\footnote{\url{http://www.astro.umontreal.ca/~bergeron/CoolingModels}}. Bivariate cubic-spline interpolation of this model grid provides absolute SDSS magnitudes at a given $\log g$ and $T_{\rm eff}$. These are then reddened for a given value of ${\rm E}(g-i)$ using the extinction law of \cite{cardelli89} and converted to apparent magnitudes for a given distance, $d$. The posterior probability distributions for the parameter set $\{\log g,\, T_{\rm eff},\, {\rm E}(g-i),\, d\}$ are estimated using the same MCMC procedure described above. We use our constraints on $\log g$ and $T_{\rm eff}$ from the Balmer-line fits as priors on these parameters. A uniform prior for ${\rm E}(g-i)$ was adopted, with values up to the galactic extinction along this line of sight (B13) allowed. A uniform prior was used for the distance. We find ${\rm E}(g-i) = 0.05 \pm 0.01$ and $d = 190 \pm 8$\,pc. Our estimates of $\log g$ and $T_{\rm eff}$ are unchanged from the values derived from the spectrum fitting. At each point in the resulting MCMC chains, the cooling models can be used to calculate a mass $M_{1}$, radius $R_{1}$ and cooling age $t_{\rm cool}$ for the white dwarf. Posterior probability distributions for each parameter are approximately Gaussian and we find most likely values for each parameter of $M_{1} = 0.53 \pm 0.03 M_{\odot}$, $R_{1} = 0.0142 \pm 0.0006 R_{\odot}$ and $t_{\rm cool} = 260\pm 20$\,Myr.

For single white dwarfs, it is possible to use an initial-final mass relation for the white-dwarf to estimate the age of the system. However, this should be considered an upper limit on the system age, since SDSS J1411 has certainly evolved through a common envelope phase, which may have ocurred prior to the AGB phase of a more massive and short-lived progenitor. Furthermore,  there is at present a significant amount of scatter between the predictions of different initial-final mass relations in the literature \citep[e.g.][]{casewell09,salaris09, catalan08}. This means it is not currently possible to constrain the age of SDSS1411 by this method.

\subsection{The $g'$-band primary eclipse of SDSS J1411}
\label{subsec:gecl}
\begin{figure}
\begin{center}
\includegraphics[width=1.0\columnwidth,trim=10 10 10 10]{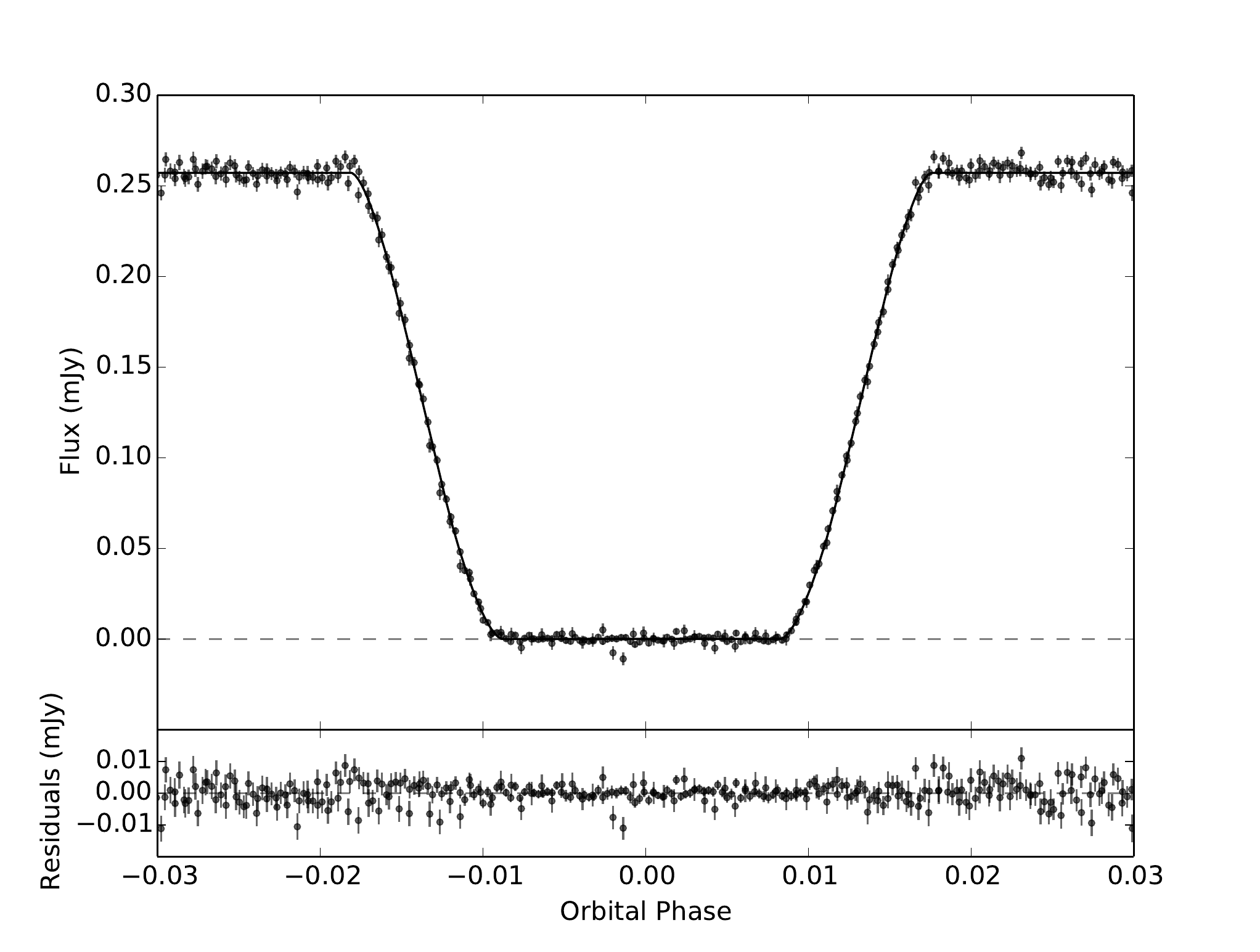}
\caption{\label{fig:gecl} The SDSS-$g'$ primary eclipse of SDSS J1411, together with the best-fit model.  A dashed line indicates zero flux. The best-fit model has $\chi^{2} = 977$ with 797 degrees of freedom.}
\end{center}
\end{figure}
The  $g'$-band primary eclipse of SDSS J1411 is shown in Figure~\ref{fig:gecl}. The ULTRACAM eclipses listed in Table~\ref{table:obs} have been phase-folded using the ephemeris of B13. The eclipse is the typical shape seen in total eclipses of a white dwarf primary. Encoded into the shape of this light curve is information about the binary inclination, and the sizes of the two components (scaled relative to the binary separation, $a$). We fit this light curve with a model which consists of the following parameters:
\begin{itemize}
\item $q = M_{1}/M_{2}$, where $M_{2}$ is the companion mass;
\item the scaled radius of the white dwarf, $R_{1}/a$;
\item the scaled radius of the companion, $R_{2}/a$;
\item the linear limb-darkening parameter for the white dwarf, $U$;
\item the inclination of the binary $i$;
\item the flux from the white dwarf outside of eclipse, $F_{1}$;
\item the flux from the companion, $F_{2}$, and
\item an offset of the eclipse centre from phase 0, $\phi_{0}$.
\end{itemize}

\begin{figure*}
\begin{center}
\includegraphics[width=2.0\columnwidth,trim=10 10 10 10]{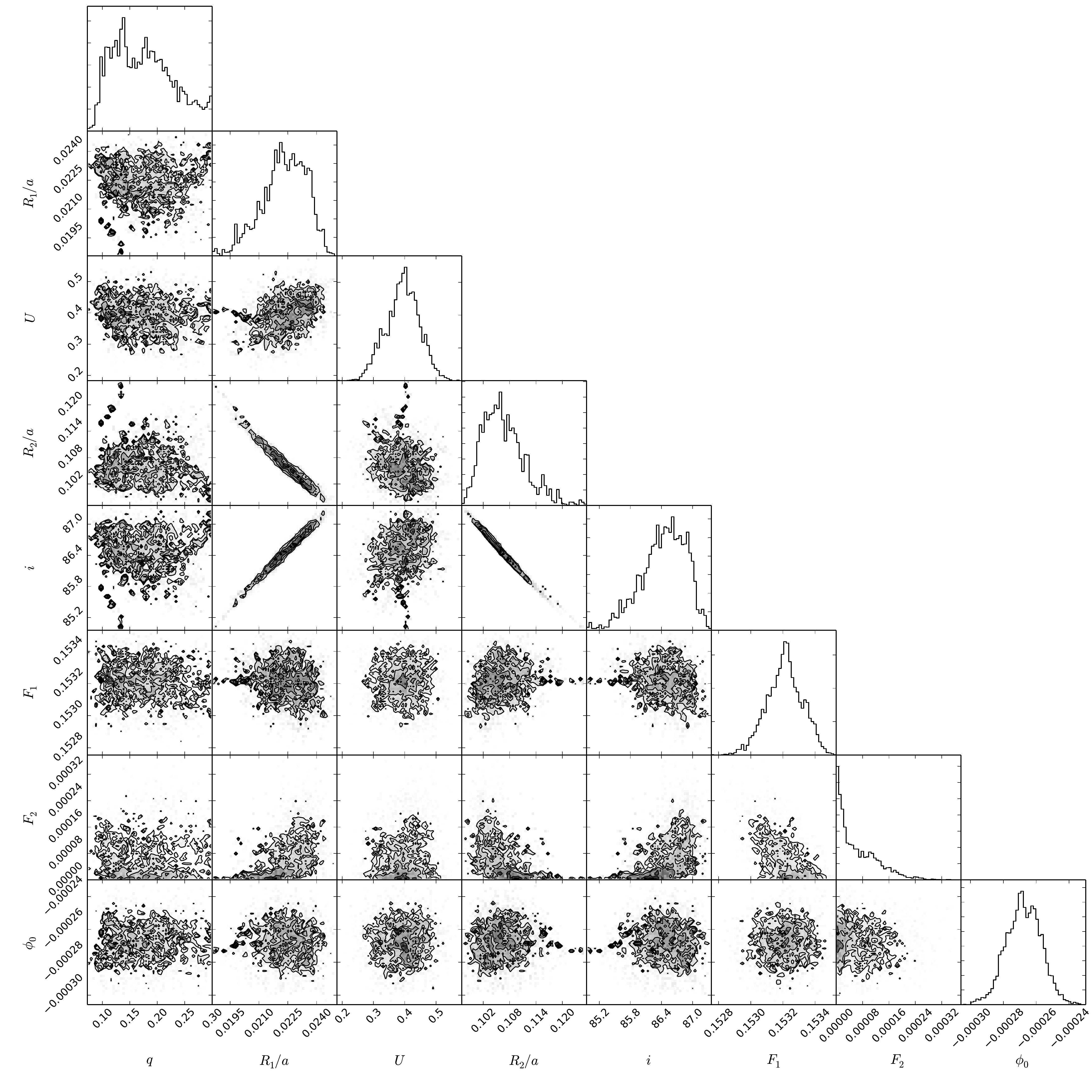}
\caption{\label{fig:geclpars} Posterior probability distributions for model parameters obtained through fitting the $g'$-band primary eclipse of SDSS J1411. See section~\protect\ref{subsec:gecl} for details of model. Greyscales and contours illustrate the joint probability distributions for each pair of parameters, whilst histograms show the marginalised probability distribution for each individual parameter. }
\end{center}
\end{figure*}

The model includes Roche distortion of the companion, and computes the visible flux from both components at each phase. Posterior probability distributions for the parameters are obtained by the same fitting procedure described in section~\ref{subsec:wd}, and are shown in Figure~\ref{fig:geclpars}. We adopted uninformative priors for most parameters. A Gaussian prior for the linear limb darkening parameter was determined from the tables of \cite{gianninas13}, using our estimates of $T_{\rm eff}$ and $\log g$. We also tried using quadratic limb darkening laws, which made no significant difference to the results.  
There are strong degeneracies in our model between three parameters; the radii of the two components $R_{1}/a$ and $R_{2}/a$ and the inclination $i$. Nevertheless, we are able to constrain these parameters, with likely values of $R_{1}/a = 0.0225 \pm 0.001$, $R_{2}/a = 0.107 \pm 0.005$ and $i = 86.5 \pm 0.4^{\circ}$. Our model fitting provides few constraints on the mass ratio $q$. We can only say that values less than 0.1 are unlikely. This is because for very low mass ratios, Roche distortion becomes important, providing a poor fit to the light curve. There is no evidence for a detection of the companion  in the $g'$-band eclipse.

\subsection{The $z'$-band primary eclipse}
\label{subsec:zecl}
\begin{figure}
\begin{center}
\includegraphics[width=1.0\columnwidth,trim=10 10 10 10]{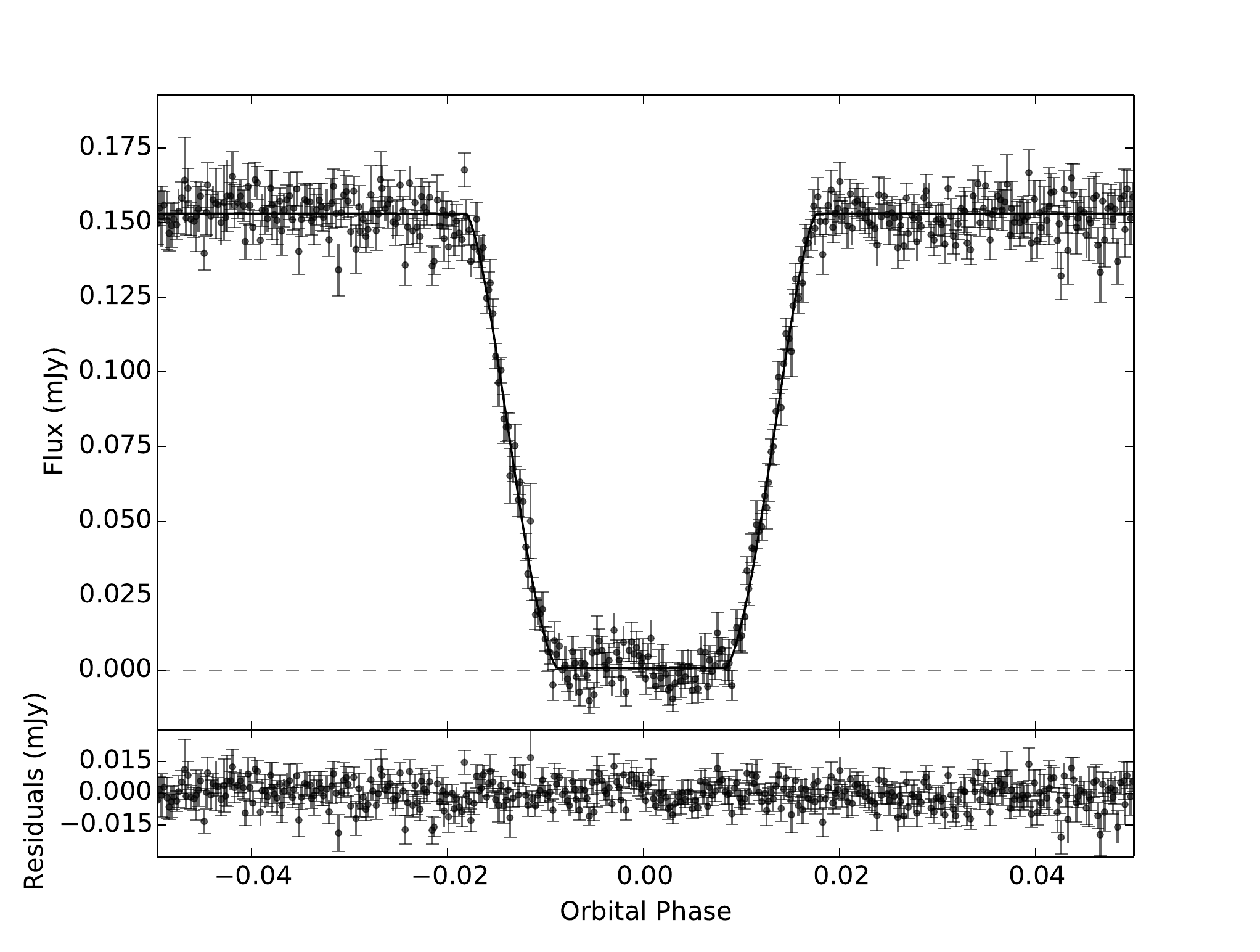}
\caption{\label{fig:zecl} The SDSS-$z'$ primary eclipse of SDSS J1411, together with the best-fit model. All data are shown in this light curve, which has been averaged into 205 bins. A dashed line indicates zero flux. The best-fit model has $\chi^{2} = 3352$ with 2481 degrees of freedom.}
\end{center}
\end{figure}
\begin{figure*}
\begin{center}
\includegraphics[width=2.0\columnwidth,trim=10 10 10 10]{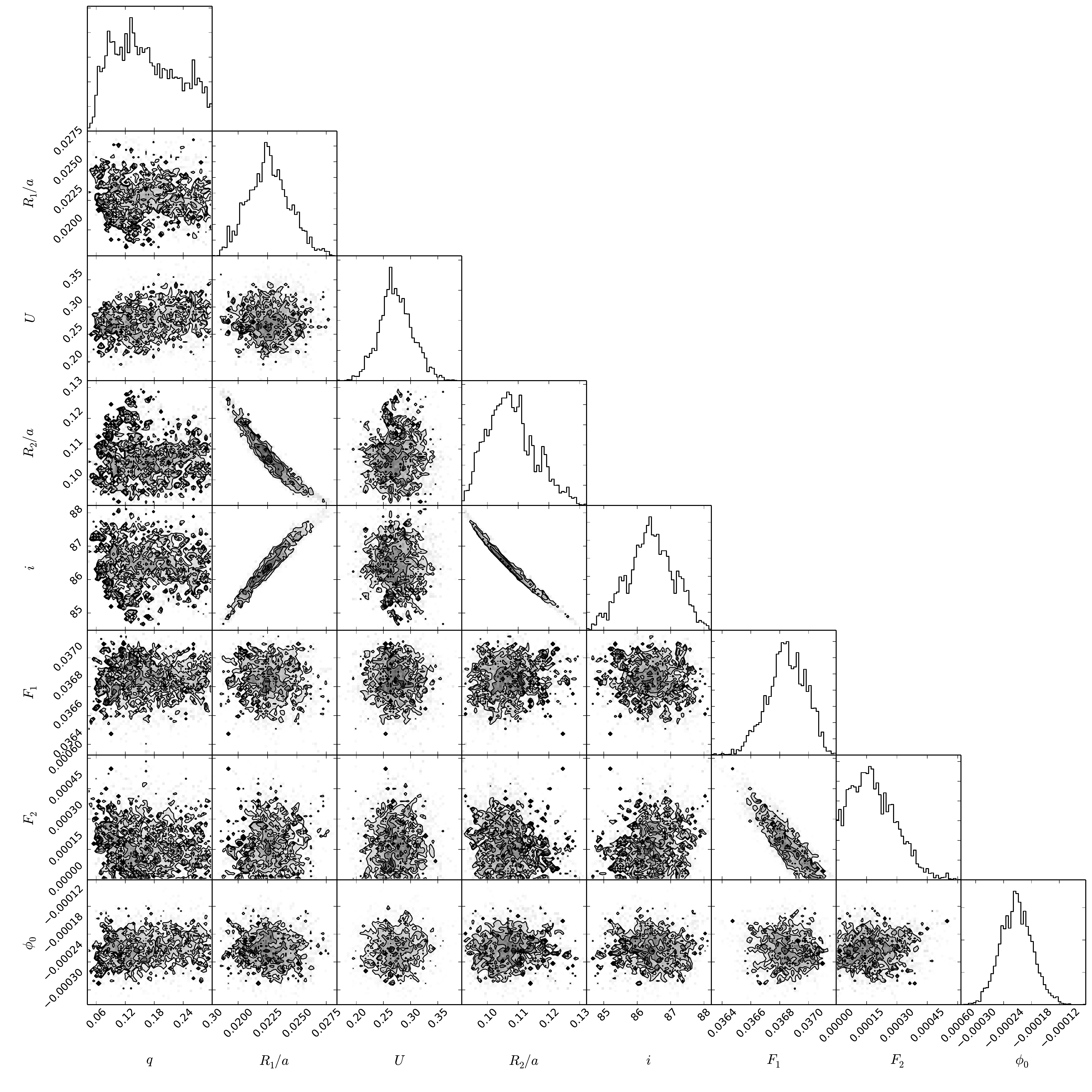}
\caption{\label{fig:zeclpars} Posterior probability distributions for model parameters obtained through fitting the $z'$-band primary eclipse of SDSS J1411. See section~\protect\ref{subsec:zecl} for details of model. Greyscales and contours illustrate the joint probability distributions for each pair of parameters, whilst histograms show the marginalised probability distribution for each individual parameter. }
\end{center}
\end{figure*}

We fit the same model described in section~\ref{subsec:gecl} to the $z'$-band eclipses, using the same priors as in the $g'$-band eclipse fitting. The posterior probability distributions for all parameters are shown in Figure~\ref{fig:zeclpars}, and the most likely model is shown in Figure~\ref{fig:zecl}.
The posterior probability distribution implies a most likely companion flux of $F_{2} = 7.5 \times 10^{-4}$\,mJy, but the distribution is also consistent with zero flux from the companion. We therefore do not claim to detect the companion at mid-eclipse, but set a 99 percent upper limit to the companion flux of $F_{2} < 1.9 \times 10^{-3}$\,mJy. We can combine this with the estimates of the distance and reddening to SDSS J1411 obtained in section~\ref{subsec:wd} to set a 99 percent limit on the {\em absolute} $z'$-band magnitude of the companion of $z'_{2} > 16.6$.  Since this limit is derived during mid-eclipse, it 
is relatively unaffected by irradiation from the white dwarf (see section~\ref{subsec:nir}).

\begin{figure}
\begin{center}
\includegraphics[width=1.0\columnwidth,trim=10 10 10 10]{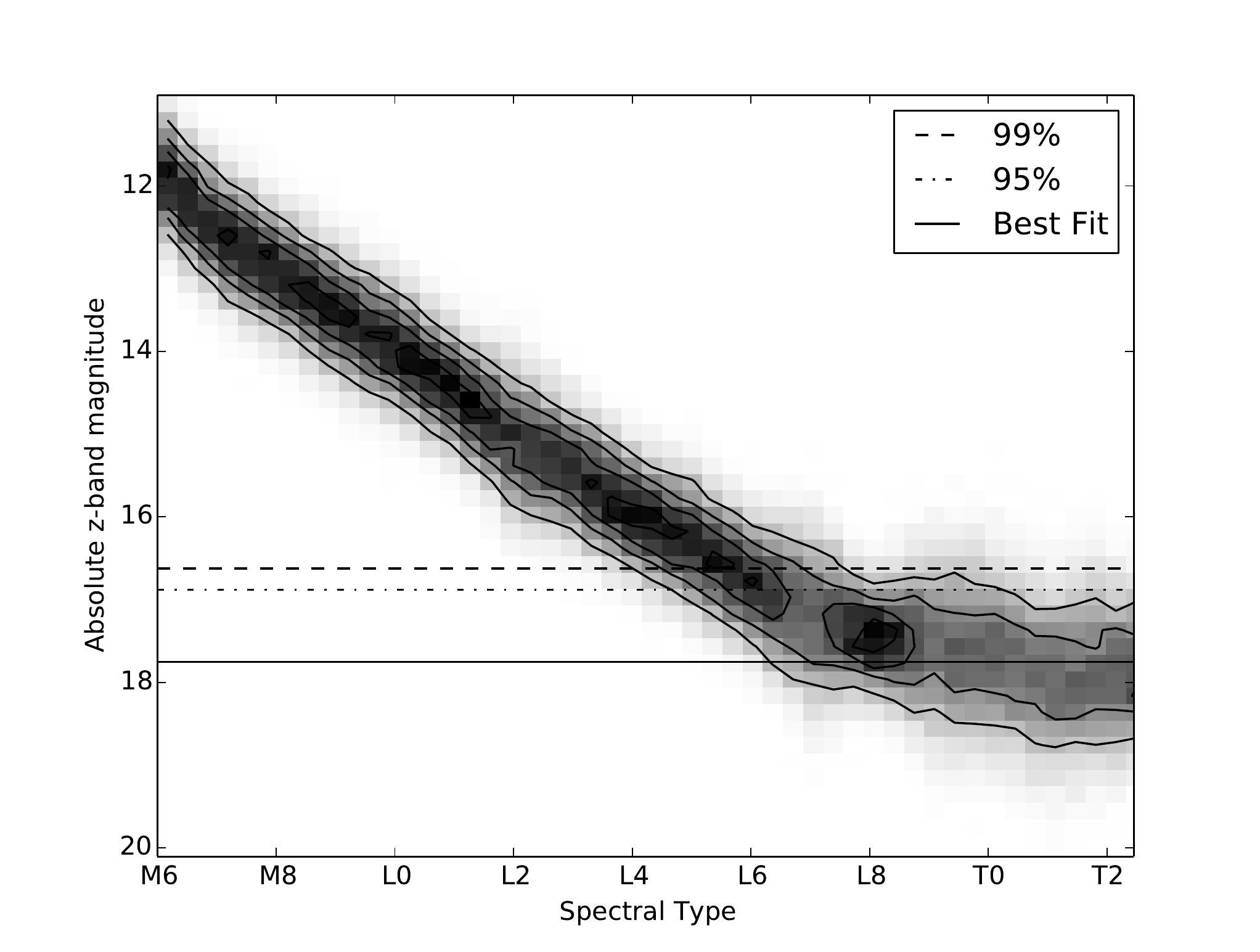}
\caption{\label{fig:zlim} Comparison of the absolute $z'$-band magnitude of the companion in SDSS J1411 with empirical estimates of the absolute $z'$-band magnitude of low mass stars and brown dwarfs. The greyscale shows the expected distribution of absolute $z'$-band magnitudes (see section~\protect\ref{subsec:zecl} for details). The solid line shows the most likely absolute $z'$-band magnitude for the companion in SDSS J1411, whilst dot-dashed and dashed lines show 95 percent and 99 percent limits, respectively.}
\end{center}
\end{figure}

We can compare these limits to empirical estimates of the absolute $z'$-band magnitudes of low-mass stars and brown dwarfs in order to estimate the spectral type of the companion. Absolute $J$-band magnitudes for field brown dwarfs from \cite{dupuy12} were combined with the observed SDSS and infrared colours from \cite{hawley02} to compute the expected absolute $z'$-band magnitudes of field objects. We took account of the scatter in all measured quantities to provide a range of expected absolute $z'$-band magnitudes at each spectral type. The results are shown in Figure~\ref{fig:zlim}. The lack of a robust detection of companion in the $z'$-band at mid-eclipse sets a 99 percent upper limit to the companion's spectral type of L5, and indicates a most likely spectral type of L8 or later. This estimate is entirely model independent, and rules out a main sequence companion in SDSS J1411.

\subsection{Eclipse Times}
\label{subsec:times}
Mid-eclipse times for each eclipse were determined by fitting the model described in section~\ref{subsec:zecl} to each $z'$-band eclipse individually. All parameters were held fixed at the best fit values derived from the average $z'$ lightcurve, with the exception of the phase offset $\phi_0$. Mid-eclipse times are shown in Table~\ref{table:obs}. Variations in the times of mid-eclipse have been used to infer the existence of planets in white dwarf binaries \cite[see][for examples]{beuermann10,marsh14}; our eclipse times show no deviation from the linear ephemeris of B13.

\subsection{NIR photometry}
\label{subsec:nir}
\begin{figure*}
\begin{center}
\includegraphics[width=2.0\columnwidth,trim=10 10 10 10]{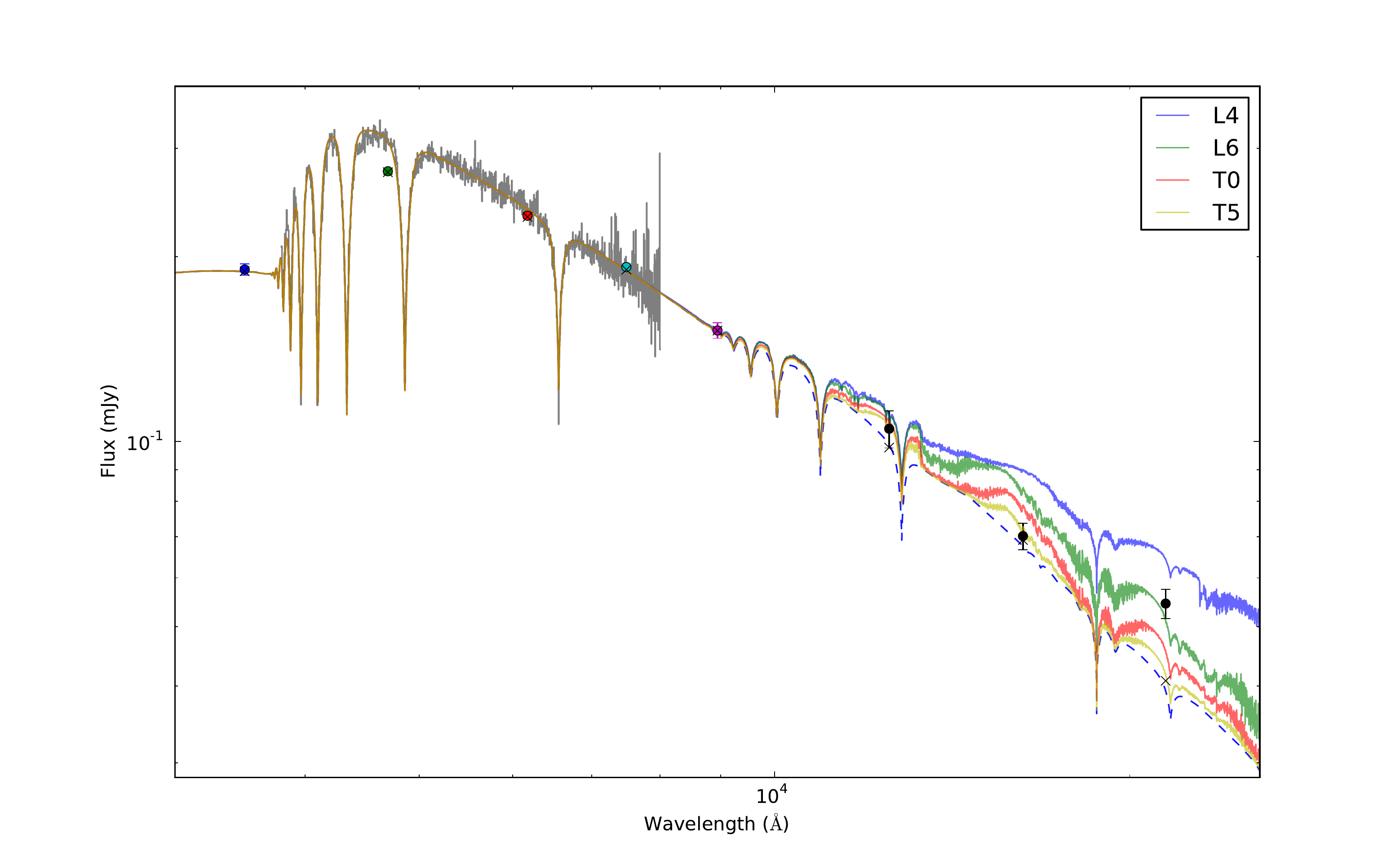}
\caption{\label{fig:excess} Comparison of selected white dwarf + brown dwarf models to the spectral energy distribution (SED) of SDSS J1411 (see section~\protect\ref{subsec:nir} for details). Combined white dwarf + brown dwarf models are shown from top to bottom with spectral types L4, L6, T0 and T5. The SDSS spectrum is shown in grey, whilst coloured points show the SDSS photometry. Our best fitting white dwarf model to the SDSS photometry and X-Shooter spectra is shown as a dashed blue line, whilst crosses mark the synthetic photometry computed by folding this model through the appropriate filter responses.   The SDSS spectrum has been scaled by a fifth order polynomial to account for uncertainties in flux calibration. Black points show the WHT/LIRIS JHK photometry.}
\end{center}
\end{figure*}

\begin{figure}
\begin{center}
\includegraphics[width=1.0\columnwidth,trim=10 10 10 10]{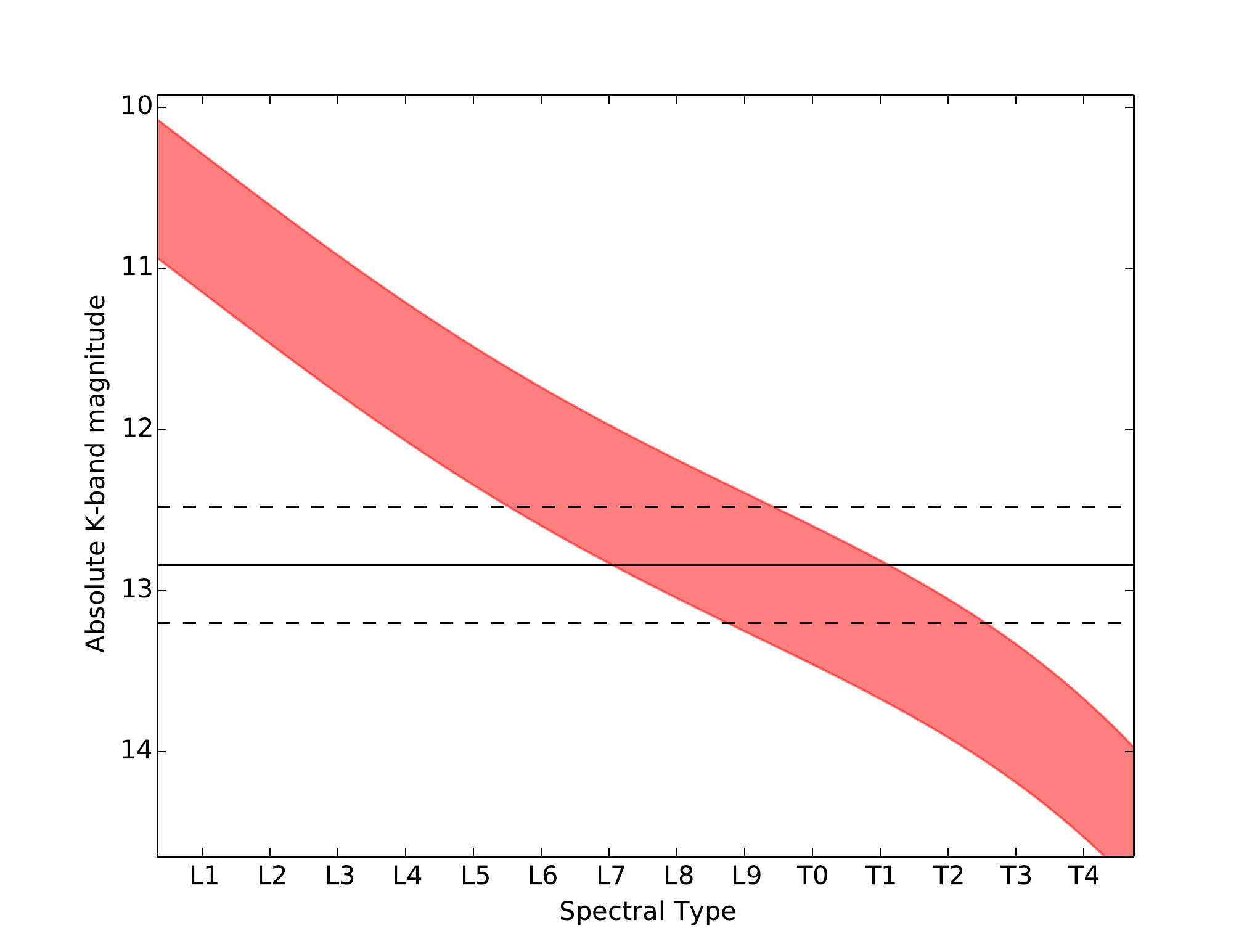}
\caption{\label{fig:klim} Comparison of the absolute $K_s$-band magnitude of the companion in SDSS J1411 with empirical estimates of the absolute $K_s$-band magnitude of low mass stars and brown dwarfs. The red shaded area shows the expected distribution of absolute $K_s$-band magnitudes (see section~\protect\ref{subsec:nir} for details). The black solid line shows the most likely absolute $K_s$-band magnitude for the companion in SDSS J1411, whilst black dashed lines show the 1$\sigma$ limits. }
\end{center}
\end{figure}

We compare the WHT/LIRIS near-infrared (NIR) photometry to the best-fitting white dwarf model in Figure~\ref{fig:excess}. The blue dashed line shows the white dwarf model fit. It is clear that a NIR excess is present in SDSS J1411. Whilst this excess is not significant in the $J$- and $H$-bands, it is highly significant in the $K_{s}$-band. We calculated the $K_{s}$ excess, accounting for uncertainties in the NIR photometry and our white dwarf model fit to give a $K_{s}$ excess of $0.014 \pm 0.004$\, mJy. Combining this estimate with the distance and extinction from the white dwarf fitting implies an absolute $K_{s}$-band magnitude of $K_{s} = 12.9 \pm 0.3$. This value is compared to the absolute $K_{s}$-band magnitudes of field brown dwarfs and low-mass stars from \cite{dupuy12} in Figure~\ref{fig:klim}. Taking into account the uncertainty in our measurements, and the scatter around the mean absolute magnitude measured by \cite{dupuy12} at each spectral type, implies a spectral type in the range L7--T1, and rules out spectral types earlier than L6, consistent with the limits placed by the $z'$-band eclipse. 

Also shown in Figure~\ref{fig:excess} are illustrative combined white dwarf/brown dwarf models produced as follows. For a given spectral type, the spectral-type effective-temperature relations of \cite{stephens09} are used to calculate an effective temperature. This is used in combination with an assumed system age of five Gyr and the evolutionary models of \cite{baraffe03} to calculate a radius and $\log g$ for the companion. The radius is combined with the distance from white dwarf fitting to scale a BT-Settl model spectrum \citep{allard11} of appropriate temperature and gravity. The flux from the white dwarf and brown dwarf model spectra are added to produce a combined model spectrum. Once again, comparison of these model spectra to the NIR photometry suggests a late spectral type for the companion. Note that the NIR photometric points represent averages of the brightness over a wide range of orbital phases (see Table~\ref{table:obs} for details). Whilst the $J$ and $K_s$ magnitudes are consistent with a companion  of spectral type L6, the $H$-band measurement is too faint, even for a comparison spectral type of T0. The $H$-band measurement is consistent with a mid-T spectral type. It is quite possible that this is a result of irradiation of the companion  by the much hotter, nearby white dwarf. The average phase of the $J$ and $K_s$-band measurements are 0.66 and 0.30, respectively. Hence these measurements contain significant contribution from the irradiated face of the companion. By contrast, the $H$-band measurement is dominated by observations of the un-irradiated face of the companion, with an average phase of 0.96. Time-resolved NIR photometry on an 8-m class telescope would be useful to determine the influence of irradiation on the companion, which may introduce variability in flux of 10 percent or more \cite[e.g.][- see also section~\protect\ref{sec:discussion}]{sudarsky03,casewell13}.

\subsection{Radial Velocity of the White Dwarf}
\label{subsec:rv}
\begin{figure}
\begin{center}
\includegraphics[width=1.0\columnwidth,trim=10 10 10 10]{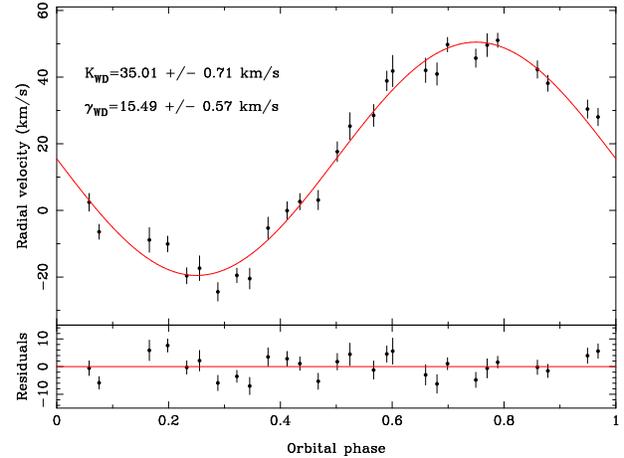}
\caption{\label{fig:rv} The radial velocity curve of the white dwarf in SDSS J1411, as measured from the NLTE core of the H$\alpha$ line.}
\end{center}
\end{figure}
The NLTE-core of the H$\alpha$ line in the X-Shooter spectrum provides a sharp feature which can be used to measure the radial velocity of the white dwarf. We measured the radial velocities of the H$\alpha$ line by simultaneously fitting all of the X-Shooter spectra. We used a combination of a straight line and Gaussians for each spectrum (including a broad Gaussian component to account for the wings of the absorption) and allowed the position of the Gaussians to change velocity according to
\begin{equation*}
V = \gamma_{1} + K_{1} \sin \phi,
\end{equation*}
where $\phi$ is the orbital phase, determined using the ephemeris of B13. The radial velocity curve of the white dwarf is shown in Figure~\ref{fig:rv}. Our fit implies a radial velocity amplitude for the white dwarf of $K_{1} = 35.0 \pm 0.7$\,km\,s$^{-1}$, and a systemic velocity for the white dwarf of $\gamma_{1} = 15.5 \pm 0.6$\,km\,s$^{-1}$. Note this systemic velocity does not represent the radial velocity of the binary centre of mass, due to the gravitational redshift of the white dwarf. We use the constraints on the white dwarf mass and radius from section~\ref{subsec:wd}, to obtain a radial velocity for SDSS J1411 of $\gamma = -9 \pm 3$\,km\,s$^{-1}$. Combining the radial velocity with proper motions for SDSS J1411 \citep{girven11} and our distance estimate, we find Galactic space velocities relative to the local standard of rest (LSR) of $(U,V,W) = (31 \pm 3, \, -13 \pm 3, \, 13 \pm 3)$\,km\,s$^{-1}$, with the sign of $U$ being positive  toward the Galactic anti-center. Following the method of \cite{bensby14}, we calculate {\em relative} probabilities that SDSS J1411 belongs to the thin disk, thick disk and stellar halo. SDSS J1411 is more than 60 times more likely to belong to the thin disk than the thick disk, and over 40,000 times more likely to belong to the thin disk than the halo. We conclude that SDSS J1411 is likely a member of the thin disk population.

\subsection{System Parameters}
\label{subsec:syspars}
\begin{table}
\begin{center}
\begin{tabular}{lcc}
\hline
Parameter & Most likely & 1-$\sigma$ error \\
\hline
\multicolumn{3}{|l|}{{\em (a) White dwarf parameters from SED fitting}} \\
\multicolumn{3}{|l|}{{}} \\
$T_{1}$ (K) & 13000 & 300 \\
$\log g$ & 7.86 & 0.07 \\
$M_{1}$ ($M_{\odot}$) & 0.53 & 0.03 \\
$R_{1}$ ($R_{\odot}$) & 0.0142 & 0.0006 \\
$t_{\rm cool}$ (Myr) & 260 & 20 \\
$d$ (pc) & 190 & 8 \\
${\rm E}(g-i)$ & 0.05 & 0.01 \\
\hline
\multicolumn{3}{|l|}{{\em (b) Fitted parameters from $g'$-band eclipse}} \\
\multicolumn{3}{|l|}{{}} \\
$R_{1}/a$ & 0.0225 & 0.001 \\
$R_{2}/a$ & 0.107 & 0.005 \\
$i$ $(^{\circ})$ & 86.5 & 0.4 \\ 
\hline
\multicolumn{3}{|l|}{{\em (c) From X-Shooter spectroscopy}} \\
\multicolumn{3}{|l|}{{}} \\
$K_{1}$ (km\,s$^{-1}$) & 35.0 & 0.7 \\
$\gamma_{1}$ (km\,s$^{-1}$) & 15.5 & 0.6 \\
\hline
\multicolumn{3}{|l|}{{\em (d) Derived system parameters}} \\
\multicolumn{3}{|l|}{{}} \\
$M_{2}$ $(M_{\odot})$ & 0.050 & 0.002 \\
$R_{2}$ $(R_{\odot})$ & 0.072 & 0.004 \\
$a$ $(R_{\odot})$ & 0.68 & 0.01 \\
$\gamma$ (km\,s$^{-1}$) & -9 & 3 \\
\end{tabular}
\caption{\label{table:syspars}System parameters for SDSS J1411.}
\end{center}
\end{table}
We can use the constraints from the optical eclipses and X-Shooter spectroscopy to obtain (white-dwarf model dependent) masses and radii for both components in SDSS J1411 as follows. The radial velocity of the white dwarf can be used to construct the mass function of the system
\begin{equation}
\frac{P}{2\pi G} K_{1} = \frac{M_{2}^{3}}{(M_{1}+M_{2})^{2}} \sin^{3} i.
\label{eq:mfunc}
\end{equation}
Combining equation~\ref{eq:mfunc} with the estimate of the inclination from the $g'$-band eclipse and the estimate of the white dwarf mass, allows the companion mass to be estimated. Kepler's 3$^{rd}$ Law is used to calculate the binary separation $a$ and hence convert the scaled companion radius, $R_{2}/a$,  to an absolute measurement.  The resulting system parameters are shown in Table~\ref{table:syspars}. Note that, due to redundancy in the parameter constraints, we can calculate $R_{1}$ either from the white dwarf spectrum and SED fitting, or by combining $R_{1}/a$ from the light curve fitting with the value of $a$ found above. The two values are consistent within $1\sigma$, giving us  confidence in our results. 

\section{Discussion}
\label{sec:discussion}
Both the lack of detection of the companion during the $z'$-band eclipse and the NIR photometry place strong constraints on the spectral type of the companion  in SDSS J1411. We can rule out companions with spectral types earlier than L5, and the likely spectral type is between L7 and T1. Since the end of the stellar main sequence occurs at a spectral type somewhere in the range L2--L4 \citep{dieterich14}, our spectral type rules out any possibility of a stellar mass companion, confirming SDSS J1411 as the first detached, eclipsing white dwarf/brown dwarf binary.

\begin{figure}
\begin{center}
\includegraphics[width=1.0\columnwidth,trim=10 10 10 10]{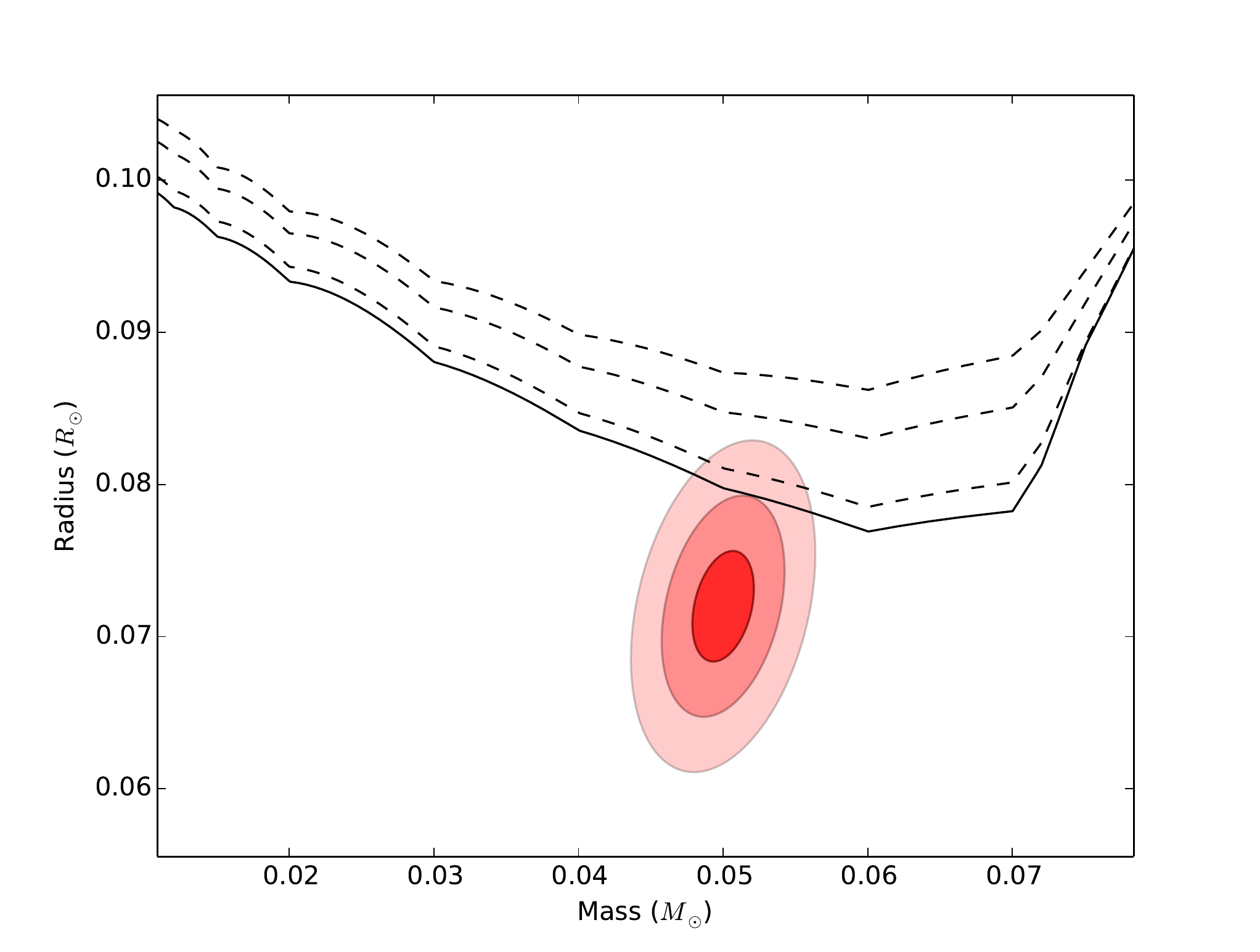}
\caption{\label{fig:mr} The mass and radius of the substellar companion in SDSS1411, compared to evolutionary models. The evolutionary models of \protect\cite{baraffe03} are shown for solar metallicity and ages of 2, 3, 6  (dashed lines) and 9 Gyr (solid line). The 1-, 2- and 3-$\sigma$ error ellipses for the mass and radius of the companion are shaded in red.}
\end{center}
\end{figure}

The mass and radius constraints on the companion are shown in Figure~\ref{fig:mr}. Also shown are evolutionary models, for solar metallicity, at a range of ages \citep{baraffe03}. If the evolutionary models are correct at these masses, then the relatively small radius for the brown dwarf  favours an age in excess of three Gyr for SDSS1411. Given the uncertainty in the system age, the measured radius is consistent with the evolutionary models. 

Since our measured companion mass and radius both depend on the white dwarf mass, which is itself model-dependent, we discuss the accuracy of the white dwarf mass estimate. There is some evidence that spectroscopically determined white dwarf masses can be in error. For example, estimates of the average white dwarf mass from gravitational redshift measurements are up to 0.06$M_{\odot}$ higher than spectroscopic determinations \citep{falcon10}. An increase of 0.06$M_{\odot}$ in our white dwarf mass would increase the companion mass and radius by 1-$\sigma$, suggesting systematic uncertainties of the same size as our formal errors. Spectroscopic white dwarf masses can also depend on whether 1-D or 3-D models are used. However, adopting the corrections suggested by \cite{tremblay13} for this dependence makes little difference to our adopted white dwarf mass. A measurement of the companion's radial velocity would allow a model independent mass to be derived, removing these uncertainties.

The inferred companion mass is interesting in the light of the measured $K_s$-band excess. The $K_s$-band data suggests an absolute magnitude of $K_{s} = 12.9 \pm 0.3$, and a spectral type between L7 and T1. By contrast, at five Gyr, the evolutionary models of \cite{baraffe03} suggest a 0.05\,$M_{\odot}$ companion would have spectral type nearer T7 and an absolute magnitude near $K_{s} = 16.2$.  As mentioned in section~\ref{subsec:nir}, it is possible that the $K_s$-band measurement is brighter than expected because it is dominated by the contribution from the irradiated face of the companion. The $z'$-band data do not suffer from this, since during primary eclipse we are viewing the un-irradiated hemisphere. Although we do not have a robust detection, our $z'$-band eclipse suggests a most likely absolute magnitude of $z' = 18.0$. By folding a BT-Settl model atmosphere of appropriate temperature through the SDSS $z'$-band filter response, we can estimate that a 0.05\,$M_{\odot}$ companion at five Gyr should have $z' = 18.2$, which is not too far from our most likely value. 

A rough estimate of the increase in temperature of the irradiated hemisphere can be given by
\begin{equation}
T^{4}_{2,irr} = T^{4}_{2} + T^{4}_{1} (R_{1}^{2}/2a^{2})(1-A_{\rm B}),
\end{equation}
where $A_{\rm B}$ is the Bond albedo. Adopting a Bond albedo of $\sim 0.5$ \citep{marley99} and a temperature for the un-irradiated hemisphere of 1000\,K (corresponding roughly to a spectral type of T7), suggests the irradiated hemisphere of SDSS 1411 should have an equilibrium temperature around 1400\,K. This would correspond to a spectral type near L8, in line with our $K_{s}$-band measurements. Therefore our data suggests that the companion  in SDSS J1411 has a significant temperature difference between hemispheres, as a result of irradiation by the white dwarf.

\section{Conclusions}
\label{sec:conclusions}
We present high time resolution optical light curves of the primary eclipse in SDSS J1411, together with time-resolved X-Shooter spectroscopy and near-infrared $JHK_{s}$ photometry.  The X-Shooter spectrum implies the white dwarf in SDSS J1411 has effective temperature $T_{1} = 13000 \pm 300$\,K,  $\log g = 7.86 \pm 0.07$ and a radial velocity of $K_{1} = 35.0 \pm 0.7$ km\,s$^{-1}$. Fitting white dwarf cooling models to the SDSS photometry with the temperature and gravity as constraints provides an estimate of the distance to SDSS J1411 of $d=190 \pm 8$\, pc, and suggests the  white dwarf has a mass of $0.53\pm0.03$\,$M_{\odot}$.

We did not detect the light from the companion during the $z'$-band eclipse. This sets an upper limit to the spectral type of the companion  of L5. Comparing the NIR photometry to the expected white dwarf flux reveals a significant $K_s$-band excess, suggesting a spectral type in the range L7--T1, although this may be dominated by emission from the warm, irradiated hemisphere of the companion. 

Combining the white dwarf mass and radial velocity with the scaled radius and inclination constraints from the eclipse light curves allows a (white-dwarf model dependent) mass and radius to be measured for the companion  of $M_{c} = 0.050 \pm 0.002$\,$M_{\odot}$ and $R_{c} = 0.072 \pm 0.004$\,$M_{\odot}$, respectively. The radius measurement is consistent with the predictions of evolutionary models, and suggests a system age in excess of three Gyr. The low companion mass is inconsistent with the inferred spectral type of L7--T1, instead predicting a spectral type nearer T5. This indicates that irradiation of the companion in SDSS J1411 could be causing a significant temperature increase, at least on one hemisphere of the companion.

\section{Acknowledgements}
The authors thank D. Koester for kindly providing his white dwarf model spectra, and T. Naylor for useful discussions. VSD, SPL and ULTRACAM are supported by STFC grant ST/J001589/1. TRM and SC are supported by STFC grant ST/L000733/1. TRM and MCP acknowledge the support of the International Exchange Scheme from the Royal Astronomical Society (grant number IE120385). SGP acknowledges financial support from FONDECYT in the form of grant number 3140585. SLC acknowledges the support of the College of Science and Engineering at the University of Leicester. SB is supported by the Foundation for Fundamental Research on Matter (FOM), which is part of the Netherlands Organisation for Scientific Research (NWO). The results presented in this paper are based on observations collected at the European Southern Observatory under the large programme ID 192.D-0270. The results presented in this paper are based on observations made with the William Herschel Telescope operated on the island of La Palma by the Isaac Newton Group in the Spanish Observatorio del Roque de los Muchachos of the Instituto de Astrofisica de Canarias. This research has made use of NASA{'}s Astrophysics Data System Bibliographic Services and the SIMBAD data base, operated at CDS, Strasbourg, France. 

\bibliography{refs,refs2,refs3}

\end{document}